\documentclass[twocolumn,showpacs,preprintnumbers,amsmath,amssymb,amsfonts,
nofootinbib,
 aps,
 prb,
floatfix,
 lengthcheck,%
]{revtex4-2}
\usepackage[english]{babel}
\usepackage[utf8]{inputenc} 
\usepackage{SIunits}
\usepackage{hyperref}
\usepackage{graphicx}
\usepackage{ulem}
\usepackage{color}
\begin{document}
\title{An~on-demand~source~of~energy-entangled~electrons~using~Levitons}
\author{B. Bertin-Johannet}
\email{bruno.bertin@cpt.univ-mrs.fr}
\author{L. Raymond}
\author{F. Ronetti}
\author{J. Rech}
\author{T. Jonckheere}
\author{B. Grémaud}
\author{T. Martin}
\affiliation{Aix Marseille Univ, Universit\'e de Toulon, CNRS, CPT, Marseille, France}

\begin{abstract}
We propose a source of purely electronic energy-entangled states implemented in a solid-state system with potential applications in quantum information protocols based on electrons. The proposed device relies on the standard tools of Electron Quantum Optics (EQO) and exploits entanglement of the Cooper pairs of a BCS superconductor. The latter is coupled via an adjustable quantum point contact to two opposite spin polarized electron wave-guides, which are driven by trains of Lorentzian pulses. This specific choice for the drive is crucial to inject purely electronic entangled-states devoid of spurious electron-hole pairs. In the Andreev regime, a perturbative calculation in the tunnel coupling confirms that entangled electrons states (EES) are generated at the output of the normal side. For arbitrary tunnel coupling and for a periodic drive, DC currents and noise (auto and cross correlations) are computed numerically using a Keldysh-Nambu-Floquet formalism.  Importantly, for a periodic drive, the production of these states can be controlled in time, thus implementing an on-demand source of entangled states. We exploit realistic experimental parameters for our device to identify its optimal functioning point.
\end{abstract}


\maketitle
Entanglement of quantum states constitutes an essential element of all quantum information protocols. Two decades ago, electronic entanglement has been proposed in mesoscopic setups and nanodevices \cite{lesovik2001,recher2001,chtchelkatchev2002,samuelsson2003}. 
BCS superconductors provide a natural source of entangled states. The Cooper pair beam splitter, a device built from a hybrid junction between a BCS superconductor and two normal metal leads, was proposed as a DC source of electronic entangled states where the two constituent electrons propagate in separate leads via a Crossed Andreev Reflection (CAR) process \cite{anantram1996,martin1996a,torres1999,lesovik2001,recher2001,sauret2004,sauret2005,chevallier2011,rech2012}. 
A manifestation of this entanglement could be found in the prediction of positive noise crossed correlations \cite{anantram1996,martin1996a,torres1999,lesovik2001,torres2001a} which were measured experimentally \cite{das2012b}. 
A definite theoretical proof of this electronic spin entanglement was subsequently proposed through  a Bell/Clauser Horne inequality violation test \cite{chtchelkatchev2002,sauret2005}. 
\begin{figure}
    \centering
    	\centering
	\includegraphics[width=0.4\textwidth]{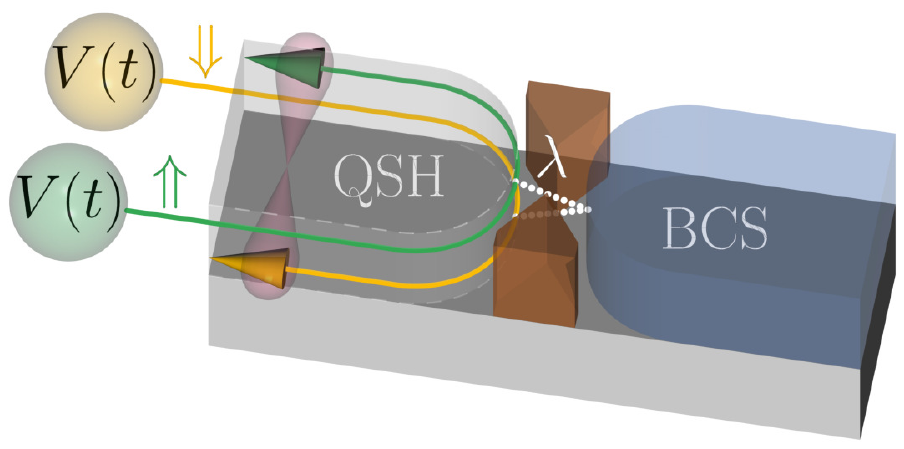}
    \caption{The setup: a superconductor (right) is tunnel coupled to a Quantum spin Hall bar with two opposite edge spin channels via an adjustable QPC. Both channels are driven by the same train of Lorentzian voltage pulses $V(t)$. The shaded area covering the channels (downstream from the injection point of the two electrons ejected from  the superconductor) represent the energy-entangled electrons which are generated on the normal side (left).}
    \label{fig1}
\end{figure}

A lot of activity has been devoted in the last decades to Electronic Quantum Optics (EQO) which reproduces quantum optics scenarios \cite{brown1956a,hong1987a} in condensed matter systems. EQO has been initially envisioned when electronic interactions are minimized \cite{dubois2013a, dubois2013b}, but nowadays EQO is also applied to strongly correlated systems such as the Fractional Quantum Hall effect \cite{rech2017a} and hybrid superconducting devices \cite{acciai2019a,Ronetti2020,bertinjohannet2022}. 
EQO has flourished due to the availability of single electron sources working in an AC regime, such as the mesoscopic capacitor \cite{feve2007a}, or voltage tailored trains of Lorentzian wave packets  called ``Levitons'', voltage drives which inject pure electron states devoid of spurious electron-hole pairs \cite{dubois2013b}.  

Here, we combine EQO tools with entanglement schemes based on singlet (BCS) superconductivity. In solid-state devices, quantum information can be encoded in electrons propagating along ballistic channels with potential applications in quantum information protocols \cite{ekert1991,bennett1992,bennett1993,long2002}. A promising path towards quantum information devices in electron systems is represented by the so-called electron flying qubits, that have been widely investigated both in nanoscopic wires \cite{yamamoto2012,bautze2014} and in surface acoustic waves \cite{edlbauer2022, helgers2022, takada2019}.

Specifically, we propose  an on-demand (periodic) source of energy-entangled electron states (EES), based on a version of the Cooper pair beam splitter: a BCS superconductor connected via an adjustable quantum point contact (QPC) to opposite spin polarized channels (Fig.~\ref{fig1}), illustrated  here as the two edge states of a quantum spin Hall bar. 
Cooper pairs are expelled from the superconductor by applying an AC Lorentzian voltage drive to the spin channels at the input and their two constituent electrons are transferred separately to the two opposite spin channels at the output as EES at each period. 
Spin entanglement is ruled out because of spin projection on these channels. 

In order to illustrate the dual nature -- spin v.s. energy -- of entanglement originating from a BCS superconductor, we write the $\pm \vec{k}$ component of the BCS wave function:
\begin{equation}
\vert \Psi_{BCS}^{(\pm \vec{k})}\rangle  \sim u_kv_k\left(c_{\vec{k}\uparrow}^\dagger c_{-\vec{k}\downarrow}^\dagger  
- c_{\vec{k}\downarrow}^\dagger c_{-\vec{k}\uparrow}^\dagger \right)\vert 0\rangle ~,
\end{equation}
where $u_k$ ($v_k$) are BCS parameters \cite{tinkham2004}. 
Strictly speaking, in the device of Fig. 1, energy -- rather than momentum -- entanglement can  be preserved for electrons tunneling to spin polarized leads~\cite{lesovik2001}. 
This non-local energy entanglement was justified by a Bell-Clauser Horne inequality \cite{bell1966} violation test in Ref.~\cite{bayandin2006}.

Recently in Ref.~\cite{bertinjohannet2022} we studied a hybrid normal metal/superconductor device where a DC and AC voltage drive was applied on the normal lead, with the goal to study minimal noise: under which conditions can electron wave packets be sent impinging on the superconductor without unwanted electron-hole pairs ? In the small superconducting gap limit $\Delta\ll \Omega$ ($\Omega$ is the frequency of the drive) where the transport process is dominated by quasiparticle transfer, integer Levitons (Lorentzian voltage pulses) lead to minimal excess noise (the difference between the period averaged noise and the equivalent DC noise) \cite{keeling2006a,bertinjohannet2022}. 
In the large gap Andreev limit $\Delta\gg \Omega$, Levitons with half-integer charge lead to minimal noise \cite{vanevic2016a,bertinjohannet2022}. 
A side result was the proposal for an on-demand Cooper pair source operating in an AC regime. 
Here, we detail the theoretical proposal of an optimized source of EES based on Levitons.

As the two constituent electrons of a Cooper pair have opposite spins, we connect the superconductor to two wave-guides with opposite spin polarization. This can be achieved by using two half metals, whose separation needs to be smaller than the superconducting coherence length (see Ref.~\cite{deutscher2000}). Here, we choose to connect the superconductor to a Quantum Spin Hall (QSH) bar (Fig. \ref{fig1}). QSH bars are two dimensional topological insulators which (thanks to spin-momentum locking) bear on their edges two spin channels which propagate with opposite chiralities. Direct Andreev reflection (rather than CAR) splits the two constituent electrons. The two spin dependent excitations which are generated are thus naturally delocalized.

We motivate our proposal by a perturbative calculation in the Andreev regime of the final quantum state propagating in the QSH channels when we apply a {\it single} Lorentzian voltage pulse to both of the spin-polarized channels
of the QSH lead.
The two pulses are assumed to be synchronised both in time and amplitude between these channels, the more general case is considered in another work \cite{bertin_private1}.
This implies that the charge of the output state is twice the charge excited by the voltage pulse.
In order to obtain a final state with a charge $2e$, corresponding to the 
emission of a single Cooper pair, we thus  excite a charge $e/2$ on each
spin-polarized lead. This is obtained by using the pulse:
\begin{equation}
   e V(t)=-\frac{1}{2 \pi W} \; \frac{1}{1+t^2/W^2}
\end{equation}
with a width $W \ll \Delta^{-1}$ in order to be in the Andreev regime.


Using second order time-dependent perturbation theory (with the Hamiltonian specified below), we computed  the quantum state generated at large times by the application of $V(t)$~\cite{bertin_private2}:
\begin{widetext}
\begin{equation}\label{eq:lev1}
    \left\lvert\mathcal{F}\right\rangle=i\frac{\lambda^2\sqrt{W}}{2\sqrt{2}\pi^2}\int_{-\infty}^{\infty}\mathrm{d}\varepsilon\, \varphi_T(2\varepsilon)\int_{-\varepsilon}^{\varepsilon}\mathrm{d}E c_{k(\varepsilon+E),\uparrow}^\dagger c_{k(\varepsilon -E),\downarrow}^\dagger\left\lvert F_\uparrow\right\rangle\otimes\left\lvert F_\downarrow\right\rangle\otimes\left\lvert\Psi_\text{BCS}\right\rangle\, ,
\end{equation}
\end{widetext}
where $\lambda$ is the tunneling amplitude between leads while $\left\lvert F_{\uparrow}\right\rangle$ and $\left\lvert F_{\downarrow}\right\rangle$ are the 
Fermi seas for the spin-polarized 
channels at equilibrium with BCS. For convenience, we set the density of states $\nu_0$ of the superconducting lead at its chemical potential such that $\pi \nu_0 =1$. 
We also introduced 
\begin{equation}
    \varphi_T(\varepsilon)=\sqrt{2W}e^{-W\varepsilon}H(\varepsilon)\, ,
\end{equation}
which is the wave function of a single Leviton~\cite{keeling2006a,Grenier2013,Glattli2016,Ronetti2018}, where $H(\epsilon)$ is the Heaviside distribution.

We stress that this state is energy-entangled, i.e., one cannot factorize it into a product of two states acting separately on the Fermi seas of the leads.
Moreover, this final state is purely electronic, i.e. devoid of additional hole-like excitations or electron-hole pairs. 
This can be achieved only by pulses with Lorentzian shape carrying a half-integer charge (or an integer multiple of it). 
An extension of this single pulse perturbative treatment to a periodic train of Lorentzian pulses will be presented elsewhere.~\cite{bertin_private2}

It seems reasonable to speculate that, in the Andreev regime, our results can be extended to the non-perturbative regime by replacing $\lambda^2 \rightarrow \lambda^2/\left(1+\lambda^4\right)$.  This conjecture is based on the fact that the vanishing of excess-noise for quantized Lorentzian pulse is indeed obtained in the non-perturbative regime of tunneling.  

We focus on the 3 terminal device represented in Fig.~\ref{fig1}. The formalism of Ref.~\cite{bertinjohannet2022} is thus naturally extended to a multi-lead setup, with an arbitrary number of (spin polarized) normal leads.
The leads are described in equilibrium by the Hamiltonians 
\begin{align}
&H_{\text{N}} = \sum_{\sigma=\uparrow,\downarrow} H_{0,\sigma,\text{N}}\nonumber\\
&H_{\text{S}} =\sum_{\sigma=\uparrow,\downarrow} H_{0,\sigma,\text{S}}+\Delta\sum_i\left(c_{i,\text{S},\downarrow}^\dagger c_{i,\text{S},\uparrow}^\dagger +c_{i,\text{S},\downarrow}^{\phantom{\dagger}}c_{i,\text{S},\uparrow}^{\phantom{\dagger}}\right),
\end{align}
where $H_{0,\sigma,j}$ is the kinetic part of the Hamiltonian of lead $j$ (with spin $\sigma$), $i$ labels the various sites of these leads, $\Delta$ is the superconducting gap and the chemical potential is set to zero, $c_{i,S,\sigma}^{\phantom{\dagger}}$ is the annihilation operator for electrons in the superconducting lead.
The tunnel Hamiltonian is defined as
\begin{equation}\label{CurrentHamApp}
    H_T(t)=\sum_{\substack{j=N,S\\j'=N,S}}\sum_{\sigma}\lambda_{j,j'}e^{i\phi_{j,j'}(t)/2}c_{e_{jj'},j,\sigma}^\dagger c_{e_{j',j},j',\sigma}^{\phantom{\dagger}}+\text{H.c.}\, ,
\end{equation}
where $\lambda_{j,j'}$ is the tunneling amplitude from lead $j$ to lead $j'$, $c_{i,j,\sigma}^{\phantom{\dagger}}$ is the annihilation operator for electrons at site $i$ and with spin $\sigma$ on the lead $j$, $e_{j,j'}$ denotes the site of lead $j$ from which tunneling to lead $j'$ occurs, and $\phi_{jj'}(t) = e\int_{-\infty}^{t} dt'\, V_{jj'}(t')$ is the time-dependent phase difference between the leads which accounts for the drive-induced voltage difference $V_{jj'}(t)$ between lead $j'$ and lead $j$.
One can introduce the tunnel matrix in lead space $W_{jj'}=\lambda_{jj'}\sigma_z e^{\sigma_z i\phi_{jj'}(t)/2}$ for leads carrying both spin orientation. 
For the device of Fig. \ref{fig1}, $\lambda_{\uparrow\downarrow}=0$, the superconducting lead only is connected to two spin-polarized chiral edge states of the QSH, so that $j$ is a spin index for the QSH. This allows us to replace the $\sigma_z$ prefactor in $W_{Sj'}$ by $(\sigma_z\pm \sigma_0)/2$ (for $j'=\uparrow,\downarrow$). We focus on $\lambda_{S \uparrow} = \lambda_{S \downarrow} = \lambda$ (as implicitly assumed in Eq.~\eqref{eq:lev1}). 

Writing the Hamiltonian as $H=\sum_j H_j + H_{\text{Tun}}$, the current operator from lead $j$ is given by
\begin{equation}\label{Current}
I_j(t)=\sum_{j'}i\psi_j^\dagger(t) \sigma_z W_{jj'}(t) \psi_{j'}(t) + \text{H.c.},
\end{equation}
where we introduced the standard Nambu spinor notation $\psi_j$ for the electron operators.
The real time irreducible noise correlator between lead $j$ and $l$ is defined as
\begin{equation}
	\tilde{S}_{jl}(t,t')=\left\langle I_j \left(t+t' \right) I_l \left( t \right) \right\rangle 
	     - \left\langle I_j \left( t+t' \right) \right\rangle
	       \left\langle I_l \left( t \right) \right\rangle \, .
\end{equation}
It is computed from the Keldysh Green's function $G^{\pm\mp}_{jj'}(t,t')= -i \left\langle {\mathcal T}_K\psi_j(t^{\pm})  \psi_{j'}^\dagger(t^{'\mp}) \right\rangle$, where ${\mathcal T}_K$ denotes Keldysh ordering, and $\pm$ superscripts stand for the branch on the contour. 
We look at $S_{jl}(t)=\int_{\-\infty}^{+\infty} dt' \tilde{S}_{jl}(t,t')$.
Because the Hamiltonian is quadratic one uses Wick' theorem to cast the noise as a product of single particle Green's functions:
\begin{widetext}
\begin{equation}\label{Noise1leadfinal}
\begin{aligned}
        S_{jl}(t)=-\frac{e^2}{\hbar}\sum_{j'l'}\int dt'\text{Tr}_{\text{N}}\big[&\sigma_z W_{jj'}(t)G_{j'l}^{-+}(t,t') \sigma_z W_{ll'}(t')G_{l'j}^{+-}(t',t)
    +\sigma_z W_{j'j}(t) G_{jl'}^{-+}(t,t')\sigma_z W_{l'l}(t')G_{lj'}^{+-}(t',t)\\
        -&\sigma_z W_{j'j}(t)G_{jl}^{-+}(t,t') \sigma_z W_{ll'}(t')G_{l'j'}^{+-}(t',t)
        -\sigma_z W_{jj'}(t) G_{j'l'}^{-+}(t,t')\sigma_z W_{l'l}(t')G_{lj}^{+-}(t',t)
        \big]\, .
\end{aligned}
\end{equation}
\end{widetext}


The periodic voltage is defined as $V(t) = V_{\text{DC}} + V_{\text{AC}}(t)$, where $V_{\text{AC}}(t)$ averages to zero on one period $T=2\pi / \Omega$ of the periodic drive. 
The injected charge per period and per spin is then given by $q = \frac{e V_{\text{DC}}}{\Omega}$. The Floquet coefficients are defined as $\exp[-i \phi(t) ]= \sum_l p_{l} e^{-i l \Omega t}$ and the Floquet weights as $P_l = \left\lvert p_l\right\rvert^2$.

The Green's functions and Dyson equations of the system adopt a double Fourier representation labeled Keldysh-Nambu-Floquet formalism which has been used profusely in Refs.~\cite{jonckheere2013,jacquet2020,bertinjohannet2022}. Fundamentally, it amounts to recasting the two-times Green's functions using two frequencies then using the periodicity in the drive frequency $\Omega$ to our advantage in order to express the correlation functions in terms of a single frequency and two harmonic indices so that the Green's function $G (\omega + n \Omega, \omega + m \Omega)$ is converted into a matrix function $G_{nm} (\omega)$.

Defining the advanced/retarded self-energy entering the Dyson equation as the matrix in Nambu-lead-harmonics space $\Sigma(t,t')=\delta(t-t')W(t)$ it can be shown that its lead-harmonics components yield
\begin{equation}
	\Sigma_{jj',mn}(\omega)=\lambda_{jj'}\begin{pmatrix}
		p_{jj',n-m} & 0        \\
		0       & -p^*_{jj',m-n}
	\end{pmatrix}\, .
\end{equation}

The time-averaged noise can then be written as the energy integral of a Nambu-harmonics trace, allowing to obtain the zero-frequency period averaged noise (PAN) $\overline{\left\langle S_{jl}\right\rangle}\equiv \int_{-T/2}^{T/2}\frac{\mathrm{d}t}{T}\left\langle S_{jl} \left(t\right) + S_{lj} \left(t\right)  \right\rangle$.

The Dyson equation in Nambu-harmonics space were solved in~\cite{bertinjohannet2022}, and the extension to the multi-lead case is straightforward.
We define the total excess noise as
\begin{equation}
    S_\text{exc} = \left. \overline{\left\langle S_T\right\rangle}\right|_{DC+AC}  - \left. \overline{\left\langle S_T\right\rangle}\right|_{\text{DC}} \, .
\label{eq:ExcessNoise}
\end{equation}
where $\overline{\langle S_T\rangle} \equiv \sum_{\sigma ,\sigma'=\uparrow,\downarrow} \overline{\langle S_{\sigma\sigma'}\rangle}$ (note that because we have a unique transport channel $\overline{\langle S_{\uparrow\downarrow}\rangle}=\overline{\langle S_{\downarrow\uparrow}\rangle}=\overline{\langle S_{\uparrow\uparrow}\rangle}=\overline{\langle S_{\downarrow\downarrow}\rangle}$, see \cite{lesovik2001}) is the {\it total} noise of the source.

\begin{figure}[t]
	\includegraphics[width=0.48\textwidth]{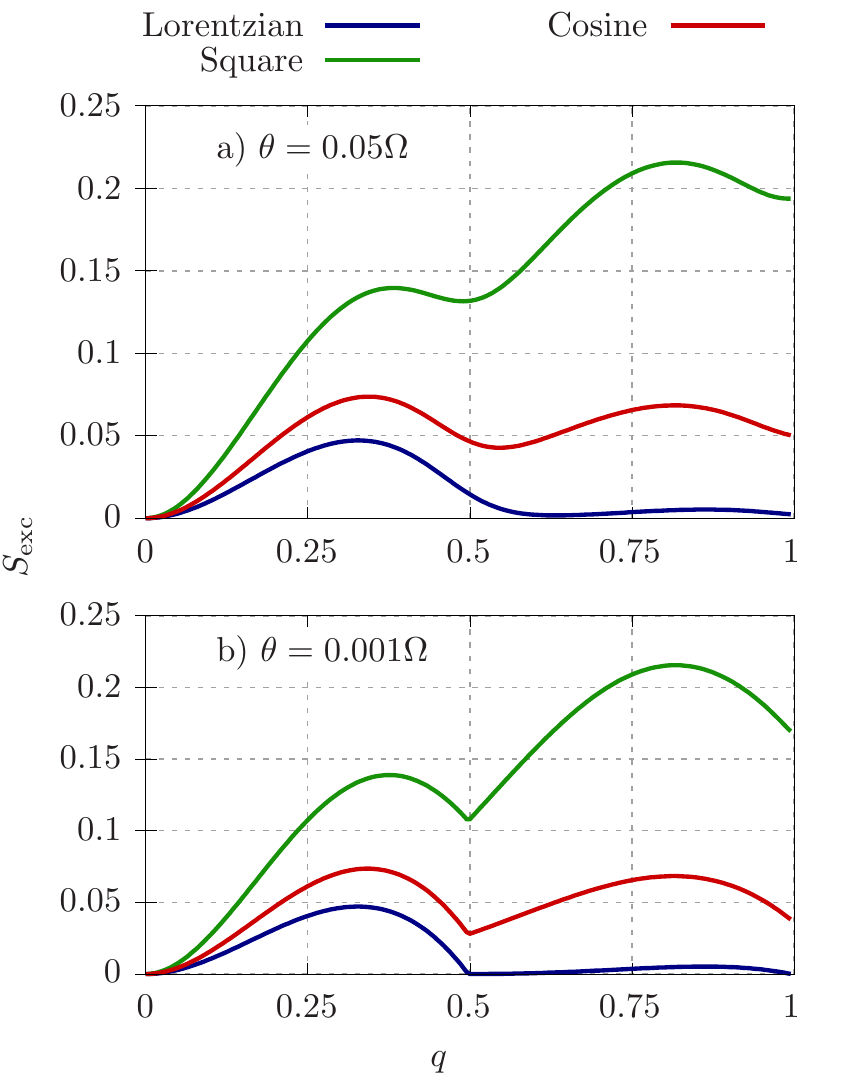}
	\caption{Excess noise for the device driven by a cosine, a periodic square and a periodic Lorentzian drive (with relative width $\eta=0.15$) as a function of the injected charge $q$ in the Andreev regime $\Omega=10^{-2}\Delta$. a) experimentally realistic temperature (see below), $\theta=5\times 10^{-2}\Omega$ and b) very low reduced temperature $\theta=10^{-3}\Omega$.} These plots are independent of the tunnel coupling (for this figure $\lambda=0.5$).
	\label{fig:fig02}
\end{figure}

In the case of a two terminal N-S device, the auto-correlations on the normal side were obtained for arbitrary $\Delta$ and $\Omega$ in Ref.~\cite{bertinjohannet2022}. The same applies for the present effective three-terminal geometry (see Fig.~ \ref{fig1}). In the pure Andreev regime $\Omega\ll\Delta$, the total noise of the source has the analytical expression: 
\begin{equation}
\begin{aligned}
		&\overline{\left\langle S_T \right\rangle}_{q} = \frac{e^2}{\pi}\bigg(4\tau_A^2\theta +2\tau_A(1-\tau_A)\\
		  &\qquad\times\sum_n  (2q+n)\Omega P_n (2q) \coth \left[ (2q+n)\frac{\Omega}{2\theta} \right]\bigg)\, ,
	\label{Eq_Noise_andreev}
\end{aligned}
\end{equation}
where $\tau_A=4\lambda^4/(1+\lambda^4)^2$. 

We observe in Fig.~\ref{fig:fig02} that with the chosen parameters, the ``exact'' numerical curves match the excess noise obtained in the Andreev regime using Eq. (\ref{Eq_Noise_andreev}). 
This excess noise is plotted both at finite (top) and zero (bottom) temperature, and minimal noise is achieved for half-integer and integer $q$ at  zero temperature, and for slightly higher values of $q$ when the temperature $\theta$ is finite. 
Periodic trains of Levitons always lead to minimal noise when compared to cosine, or square voltages, as defined in Ref.~\cite{bertinjohannet2022}:
\begin{equation}
    \begin{aligned}
        V_\text{cos}(t) &= V_{DC}(1-\cos{\Omega t})\\
        V_\text{squ}(t) &= V_{DC} \left[ 1+\text{sgn}\left(\cos{\Omega t}\right) \right] \\
        V_\text{lor}(t) &= V_{DC} \left( \frac{1}{\pi} \sum_k \frac{\eta}{\eta^2 + (t/T - k)^2 }  \right) \, .
    \end{aligned}
\end{equation}

The final step of this study is to derive a superconducting extension of Levitons in normal-metal devices~\cite{levitov1996a,dubois2013b}. The source relies on Andreev reflection (subgap regime), requiring that $\Omega,\theta\ll \Delta$. 
Borrowing from previous experiments on Levitons in metals~\cite{dubois2013b}, a realistic electron temperature is $\theta \approx \unit{10}{\milli\kelvin}$ with a drive frequency $f \approx \unit{5}{\giga\hertz}$ (this means applying voltages of the order $V_{\text{DC}} \approx\unit{10}{\micro\volt}$ to reach $q=0.5$). 
For example, niobium (with a gap $\Delta_\text{Nb}\approx \unit{1.55}{\milli\electronvolt} $) puts us well into the Andreev regime, as we have $\beta \Delta_\text{Nb} \approx 2000$ and $\frac{\Delta_\text{Nb}}{\Omega} \approx 100$. 
For aluminum ($\Delta_\text{Al} \approx \unit{0.17}{\milli\electronvolt}$), however, the situation is not as optimal, with  $\beta \Delta_\text{Al} \approx 200$ and $\frac{\Delta_\text{Al}}{\Omega} \approx 10$.

To achieve a reliable, controlled source of EES, the average charge $\langle Q \rangle$ transmitted through the junction per period should be quantized, corresponding to the Cooper  pair charge, and the excess noise should be small so as to generate primarily minimal excitation states \cite{bertinjohannet2022}:
\begin{equation}
	\langle Q \rangle = 2\pi\frac{\overline{\left\langle I \right\rangle}_{q}}{\Omega}=4 q e \tau_A\, ,
	\label{eq:avgQ}
\end{equation}
An ideal source of energy-entangled states would thus require to operate at near perfect transmission and tune the bias voltage such that the injected charge $q$ is half-integer, leading to the emission of exactly $2q\in\mathbb{N}$ Cooper pairs per period into the leads. Surprisingly, this turns out to be independent of the type of AC drive considered. However, the second property of such an ideal source, the minimization of excess noise, can only be achieved by a Leviton voltage drive, as we will show below.

The quantized periodic Lorentzian drive shows the lowest excess noise, Fig.~\ref{fig:fig02}. In the limit of zero temperature, this is demonstrated by considering the expression for the excess noise in the Andreev regime, obtained from  Eq.~\eqref{Eq_Noise_andreev},
\begin{align}
S_{\text{exc}}(q) \propto \sum_n P_n \left\lvert 2q + n \right\rvert \left[ 1 - \text{Sgn} \left( 2q + n \right)\right]\, .
\end{align}
Since this is a sum of positive terms, every single term should vanish in order to cancel the excess noise. This is obvious for $n > -2q$, but for the other contributions to also vanish, one needs $P_n= 0$ for all $n \leq -2q$ which is only satisfied by a periodic Lorentzian drive with half-integer injected charge~\cite{dubois2013b,rech2017a}. This is in agreement with the perturbative result obtained in Eq.~\eqref{eq:lev1} for single pulses.

It thus follows that the ideal source of EES corresponds to a QSH/BCS junction which operates in the Andreev regime driven by a periodic Lorentzian drive with half-integer injected charge $q$, and operating at zero temperature with perfect transmission. Unfortunately, none of these conditions can be realistically met in an actual experiment. What happens when we relax these constraints ?

\begin{figure}[t]
	\centering
	\includegraphics[width=0.48\textwidth]{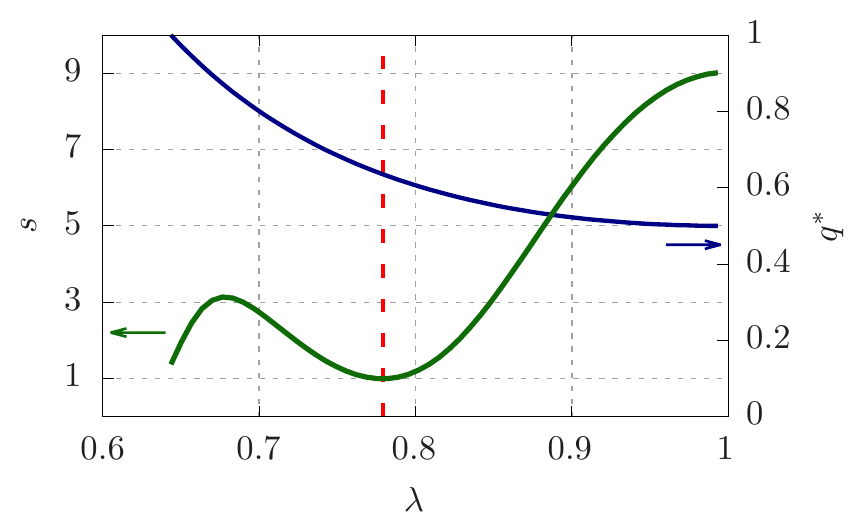}
	\caption{The total excess noise of the source  in units of the minimal noise $S_\text{min}$ (green line, left axis) and the adjusted value of the injected charge per period (blue line, right axis) as a function of the tunnel parameter $\lambda$ in the Andreev regime, $\Omega=10^{-2} \Delta$, at temperature $\theta = 5 \times 10^{-4} \Delta$. $\lambda$ covers the interval $\tau_A \in [0.5 ; 1 ]$. The dashed red line line corresponds to a transparency $\tau_A \simeq 0.79$ for which the excess noise is minimal.}
	\label{On_demand_CP}
\end{figure}

In fact, the average charge transferred per drive period is robust under variations of the electron temperature (the zero-temperature result of Eq.~\eqref{eq:avgQ} provides a very good estimate, even for the less optimal case of aluminum). This  linear in $\tau_A$ behavior suggests, however, that while working at finite temperature should hardly affect $\left\langle Q \right\rangle$, departing from perfect transmission has severe consequences, as the transmitted charge is no longer perfectly quantized. One can circumvent this issue by adjusting the injected charge to a new value $q^*$ in order to reach the optimal value $\left\langle Q \right\rangle = 2 e$ for the source of entangled Levitons.

In the Andreev regime, at $\theta=0$, the excess noise  $\propto\tau_A \left( 1 - \tau_A \right)$, which is typical of shot noise \cite{martin05}. However, the excess noise is sensitive to a nonzero temperature $\theta$. Indeed, as shown in Fig.~\ref{fig:fig02}a, the excess noise gets modified as the temperature is increased, leading to a first arch which, instead of vanishing exactly at $q=0.5$, now reaches a (nonzero) local minimum for a slightly higher value of $q$. The value $S_\text{min}$ of this local minimum, which, in the Andreev regime, only depends on the drive frequency and temperature, constitutes a good reference point to optimize the quality of the produced entangled states.
We study how the excess noise of our device varies as a function of $\lambda$ when the voltage source is operated so as to maintain a quantized value $\left\langle Q \right\rangle = 2 e$. 
In Fig.~\ref{On_demand_CP}, we show the ratio $s = \frac{S_\text{exc}}{S_\text{min}}$  as a function of $\lambda$ for an injected charge $q^* =  \frac{1}{2\tau_A}$ and $\theta = 5 \times 10^{-4} \Delta$ and $\Omega = 10^{-2} \Delta$ (we choose Niobium). Quite remarkably, there is a range of tunneling amplitudes (which are centered around $\lambda \simeq 0.78$, i.e., $\tau_A \simeq 0.79$) for which the excess noise is close to the minimum allowed for this choice of temperature and drive frequency. For $q^*\simeq 0.65$, the Lorentzian drive displays the minimal noise, as can be seen in Fig.~\ref{fig:fig02}. While a high transparency is desirable, a fully transparent junction is not the best available choice. When operated at this optimized transmission, the junction displays the minimal possible noise while transferring a $2e$ charge per period distributed on both leads on average. 

The entangled electron states produced by this source can be exploited to implement a variety of quantum protocols in solid-state devices. Among these protocols, major examples are quantum teleportation~\cite{bennett1993}, i.e the transmission of quantum information through a classical channel by exploiting entanglement, quantum key distribution~\cite{long2002}, which allows for the creation of secure cryptography~\cite{ekert1991,bennett1992,Pan2020}, or quantum dense coding~\cite{Bennett1992b}, where entangled sources are exploited to send classical bits with a reduced number of qubits.

\acknowledgements
This work received support from the French government under the France 2030 investment plan, as part of the Initiative d'Excellence d'Aix-Marseille Université- A*MIDEX. We acknowledge support from the institutes IPhU (AMX-19-IET008) and AMUtech (AMX-19-IET-01X). 

\bibliography{on_demand_entangled}
\end{document}